\documentstyle[12pt]{article}
\newcommand{\be}{\begin{equation}}
\newcommand{\ee}{\end{equation}}
\newcommand{\ba}{\begin{eqnarray}}
\newcommand{\ea}{\end{eqnarray}}
\newcommand{\bm}{\bibitem}

\begin{document}

\begin{center}
{\Large Entropy of Scalar Field in 3+1 Dimensional Constant
Curvature Black Hole Background}\\
[6mm]
K. Ghosh\footnote{E-address: kaushik@tnp.saha.ernet.in} \\
[4mm]
{\sl Saha Institute of Nuclear Physics, 1/AF, Bidhannagar, Calcutta - 700064,
India.}
\end{center}

\vspace{1.5cm}

\begin{abstract}

We consider the thermodynamics of minimally coupled massive
scalar field in 3+1 dimensional constant curvature black hole
background. The brick wall model of 't Hooft is used. When
Scharzschild like coordinates are used it is found that
apart from the usual radial brick wall cut-off parammeter
an angular cut-off parameter is required to regularize
the solution. Free energy of the scalar field is 
obtained through counting of states using the WKB
approximation. It is found that the free energy 
and the entropy are logarithmically divergent in both
the cut-off parameters.

\vspace{0.5cm}
\end{abstract}

\newpage

\section{{Introduction}}

A black hole has a horizon beyond which
no matter or information can escape. The
absence of information about the region inside the horizon
manifests itself in an entropy. A quantitative expression
for the entropy and the laws of black hole thermodynamics
were first obtained by Bekenstein [1] mainly on the basis of analogy.
Since then a lot of effort has been devoted to explain this entropy on
a statistical mechanics basis. A related issue
is the entropy of quantum fields in black hole backgrounds.
The  entropy  of  quantum  fields is obtained by various methods,
e.g.,
by tracing over local degrees of freedom inside the horizon
(geometric entropy) [2], by explicit counting of degrees of freedom
of the fields propagating outside the horizon
(entanglement entropy) [3,4,5] or by 
the Euclidean path integral [6,7].
These expressions are proportional to the area of the horizon
and constitute the first quantum correction to the gravitational
entropy. Divergences appear in the density of
the states associated to the horizon and can be absorbed in the
renormalized expression of the gravitational coupling constant [8].
To regularize these divergences 't Hooft proposed [3] that the field modes
should be cut off in the vicinity of the horizon by imposing
a brick wall cut-off. This method has  been  used  to  study  the
entropy of matter around different black hole solutions.

A new type of black hole solution has been found by
Ba{\~n}ados [9]. This solution results from an identification
of space-time points in anti-de Sitter space and represents a
higher dimensional generalization of the 2+1 dimensional BTZ black hole [10]
These are constant curvature black holes (CCBH) with a negative
cosmological constant. However the spatial topology is toroidal.
Mann and Creighton
considered the thermodynamics of the 3+1 dimensional constant curvature
black holes as a solution to the equations of general relativity
with a negative cosmological constant [11]. They used a quasilocal
foliation which is degenerate along a particular direction. The
entropy is obtained from the microcanonical action evaluated
in the Euclidean sector of the black hole space-time.
The entropy is obtained as the integral of a Noether charge
2-form on the history of the boundary of the quasilocal
surface. The entropy is unusual in the sence that it differs
from the usual expression of a quarter of the horizon surface area.
Moreover, the entropy vanishes as the boundary of the
quasilocal suface used to foliate the space-time is
pulled back to the horizon. Thus the entropy seems
to be associated with the surface of degeneracy of the
foliation used rather than the horizon surface area.

In this context it is natural to enquire about
the entropy of quantum fields defined on such backgrounds.
Thus we investigate the thermodynamical behaviour of a real
scalar field propagating on the 3+1 dimensional
constant curvature black hole using the brick wall cut-off.

\section{{Wave equation for scalar fields in
3+1 dimensional CCBH black holes:}}

The anti-de Sitter space in 3+1 dimension is defined as the
universal covering space of the hypersurface
\be
-x^{2}_{0} + x^{2}_{1} + x^{2}_{2} + x^{2}_{3} - x^{2}_{4} = - l^{2}.
\ee
The 3+1 dimensional CCBH is obtained by making identifications
in this space using a one dimensional subgroup
of its isometry group SO(2,3). In Schwarzschild - like
coordinates the metric for 3+1 dimensional CCBH is given by [9]
\be
ds^{2} = {l^{4}f^{2}(r)\over r^{2}_h}[d{\theta}^{2} - \cos^{2}{\theta}(dt/l)^{2}]
+ {dr^{2}\over f^{2}(r)} + r^{2}d{\phi}^{2}
\ee
where ${f(r) = ({{{r^{2}} - {r^{2}_h}}\over{l^{2}}})}$.
These coordinates are valid outside the horizon ($r>r_{h}$)
for $0\le{\theta}<2{\pi}$
and $0\le{\phi}<2{\pi}$. However in these coordinates only part of the
space is covered. It is clear that the foliation becomes degenerate
along the direction  ${\theta} = {\pi}/2$ and ${\theta} = 3{\pi}/2.$

The wave equation for a minimally coupled scalar field in a
curved background is
\be
{1\over\sqrt{-g}}{\partial_{\mu}}({\sqrt{-g}}{g^{\mu\nu}}{\partial_{\nu}{\psi}})
-m^{2}{\psi} = 0.
\ee
In this case it leads to
\ba
{-{l^{2}\over{\cos^{2}{\theta}}}{\partial^{2}_t}{\psi}} &+&
 {1\over{\cos{\theta}}}{\partial_{\theta}}({\cos{\theta}{\partial_{\theta}{\psi}}})
 + {{l^{4}f(r)}\over{rr^{2}_h}}
{\partial_r}[{rf^{3}(r){\partial_r}{\psi}}]\nonumber\\
 &+& {{l^{4}f^{2}(r)}\over{r^{2}r^{2}_h}}{{\partial^{2}_{\phi}}{\psi}} -
 {m^{2}l^{4}f^{2}(r)\over{r^{2}_h}}{\psi} = 0.
\ea
Since  there is no explicit time dependence
in the foliation used, we may construct
stationary state solutions.
We take the stationary state solutions to be of the
form
\be
{\psi} = K{e^{iEt}}{e^{iN{\phi}}}{P({\theta})}{R(r)}
\ee
where $N$ is an integer, $K$ a normalization constant, $E$ a real parameter.
We have, using the separation of variables,
\be
{1\over{\cos{\theta}}}{\partial_{\theta}({\cos{\theta}}{\partial_{\theta}P})}
+ {[{\nu}({{\nu}+1}) + {E^{2}l^{2}\over{\cos^{2}{\theta}}}]P} = 0,
\ee
where $\nu$ is an arbitrary complex parameter.
For the radial part, we have,
\be
{\partial_r{[rf^{3}(r){\partial_r{R}}]}} - [{N^{2}f(r)\over{r}}
+  {m^{2}rf(r)}  + {{\nu}({\nu}+1)rr^{2}_h\over{l^{4}f(r)}}]{R} =
0.
\ee

We want to calculate the entropy of the scalar field. 
For this purpose we use the WKB approximation to the radial
differential equation to obtain the radial degeneracy factor
associated with the brick wall boundary condition. Wehave to
find out the quantization condition on the parameters of the
angular equation using appropriate boundary conditions.

\section{{Calculation of entropy}}

The equation for the radial part is given by
\be
{({r^{2} - r^{2}_h}){\partial^{2}_r}{R}} + {[4r -
{{r^{2}_h}\over{r}}]}{\partial_r}{R} - {[{{N^{2}l^{2}}\over{r^{2}}} +
{m^{2}l^{2}} + {{\nu}{({\nu} + 1)}{r^{2}_h}\over{(r^{2} - r^{2}_h)}}]}{R} = 0.
\ee
The boundary conditions in this case are nontrivial. Apart from
the brick wall boundary condition $R = 0$ for $r = r_h + \epsilon$ there
is a complication arising from the fact that the adS space-time in which
the CCBH is immersed is not globally hyperbolic. The infinity in this
case is timelike. Fresh information and matter can come in
or leak out through this time-like infinity in finite coordinate time.
Moreover any initial data of compact support on a Cauchy surface
will not remain so on other space-like surfaces.  The authors of [12]
have considered in detail the
quantization of scalar fields in adS and its covering spaces. A consistent
scheme can be devised if we take the following boundary conditions:

(1) Dirichlet boundary condition:
$$
{{\lim_{r\rightarrow\infty}}{\sqrt{r}}{R}} = 0.
$$

(2) Newmann boundary condition:
$$
{{\lim_{r\rightarrow\infty}}{r^{3\over{2}}}{d\over{dr}}{({1\over{\sqrt r}}{R})}} = 0.
$$

The solution of equ.(16) for large $r$ is given by
\be
R = {1\over{{\sqrt{r^{3}}}}}{[{C_{1}{r^{\sqrt{(9+4m^{2}l^{2})\over{2}}}}} +
{C_{2}{r^{-{\sqrt{(9+4m^{2}l^{2})\over{2}}}}}}]}.
\ee
For real particles $({m^{2} = 0})$,
only the second part of the solution is acceptable
since it is in accordance with
the boundary condition at infinity.

However we are interested in the degeneracy of the field modes
near the horizon. We use the WKB approximation, i.e,
we write $R(r) = \rho(r){e^{iS(r)}}$, where $\rho(r)$ is
a slowly varying amplitude and $S(r)$ is a rapidly
varying phase factor. When substituted in equ.(8),
this gives us the following value for the $r$-dependent
wave number  

\be
k^2 = -{\big[{\nu(\nu + 1){r^2_h}\over{(r^2 - {r^2_h})^2}}
+ {{m^2}{l^2}\over{(r^2 - {r^2_h})}} + 
{{N^2}{l^2}\over{{r^2}{(r^2 - {r^2_h})}}}\big]}
\ee
We will consider only those values of $\nu$ for which
$k(r)$ is real. In the near horizon region there are
two sectors of the values of $\nu$ for which it is
possible, viz.,
\be
\nu = -{1\over2} + i\chi
\ee
where $\chi$ is a real parameter, and
\be
\nu = -\lambda(1 - \lambda)
\ee
where $\lambda$ is a real parameter with the 
range $0\leq\lambda\leq1$.

The values of $\nu$ are further restricted by the 
following semiclassical quantization condition,
\be
{{n_r}\pi} = {\int{dr {k(r,\nu,N)}}}
\ee
where $n_r$ is a nonnegative integer.This gives us 
the radial degeneracy of the mode functions.

For the first sector, we have,
\be
n_r({\chi,N}) = {1\over\pi}{\int_{r_h + \epsilon}^{L_2}dr
{\big[{{{(\chi^2 + {1\over4})}{r^2_h}\over({r^2} - {r^2_h})^2} -
{{N^2}{l^2}\over{r^2}({r^2} - {r^2_h})} -
{{m^2}{l^2}\over({r^2} - {r^2_h})}}\big]^{1\over2}}}
\ee
and for the second sector we have,
\be
n_r({\lambda,N}) = {1\over\pi}{\int_{r_h + \epsilon}^{L_1}dr
{\big[{{{\lambda(1 - \lambda)}{{r^2}_h}\over({r^2} - {r^2_h})^2} -
{{N^2}{l^2}\over{r^2}({r^2} - {r^2_h})} -
{{m^2}{l^2}\over({r^2} - {r^2_h})}}\big]^{1\over2}}}
\ee
In the above two expressions we cannot take the upper
limits of integration, i.e, ${L_1}$ and ${L_2}$ 
to be arbitrarily large. Then
the mass term will become large and the solution will no
longer remain oscillatory. This is expected from the point of
view of the discussion given below equ.(8).

To solve equation (6) and obtain quantization condition on
$E$, we put $\mu = \pm iEl$. Then equ. (6) becomes
\be
{{1\over{\cos{\theta}}}{\partial_{\theta}}({\cos{\theta}}{\partial_{\theta}P})}
+ {[{\nu}({{\nu}+1}) - {{\mu}^{2}\over{\cos^{2}{\theta}}}]P} = 0.
\ee
The solution of this equation is $ P^{\pm {\mu}}_{\nu}({\pm x})$, where
$ x = \sin{\theta}$, and
\be
{P^{\mu}_{\nu}(x)} = {1\over{{\Gamma}({1 - {\mu}})}}{{({{1+x}\over{1-x}})}^{{\mu}\over2}}
{F(-{\nu} , {{\nu} + 1}; {1-{\mu}}; {{1-x}\over2})}
\ee
with the condition that $(\nu + \mu) \neq$  an integer. Here
${F(-{\nu} , {{\nu} + 1}; {1-{\mu}}; {{1-x}\over2})}$ is the
hypergeometric function [13].
However this solution is divergent at $ x = \pm 1$ , i.e, along
$ \theta ={\pi\over 2},  {3\pi\over 2} $,  which is expected
because the foliation used is degenerate in those directions.

A consistent regularization scheme, which makes the solution
vanish at $ x = \pm (1 \mp {\eta}) $ and thus avoids the problem
associated with the degeneracy of the foliation used,
can be devised if we chose the solution to be of the following form:
\be
u = {C_1}{P^{iEl}_{\nu}(x)} + {C_2}{P^{-iEl}_{\bar\nu}(x)}
\ee
for $ {0 \leq x} $, i.e,
for $ {0 \leq {\theta} < {\pi}}$
and
\be
u = {C_3}{P^{iEl}_{\nu}(-x)} + {C_4}{P^{-iEl}_{\bar\nu}(-x)}
\ee
for $ {0 \geq x} $, i.e, for $ {{\pi} \leq {\theta} < 2{\pi}}$.

The constants $ {C_1}, {C_2}, {C_3}$ and ${C_4}$ should be chosen
in such a way that the solutions are real. Moreover the two 
forms of the solutions in the two regions 
and their first derivatives should match along ${\theta = 0}$
and ${\theta = \pi}$. However, for the first sector of the given values
of $\nu$ as determined from equ.(11), the solution (17) is
of the form of conical function [13]. The solution will
satisfy all the conditions as stated above for the 
following values of [14]  $ {C_1}, {C_2}, {C_3}$ and ${C_4}$.
$$
{C_1} = -{C_3} = {{{i{\Gamma({{3\over4} - {{i\chi}\over 2} - {{iEl}\over
2}})}
{\Gamma({{3\over4} + {{i\chi}\over 2} - {{iEl}\over 2}})}}}\over
{{2^{iEl}{\sqrt{\pi}}}}}
$$
and
$$
{C_2} = -{C_4} = {-{{i{\Gamma({{3\over4} - {{i\chi}\over 2} + {{iEl}\over
2}})}
{\Gamma({{3\over4} + {{i\chi}\over 2} + {{iEl}\over 2}})}}}\over
{{2^{-iEl}{\sqrt{\pi}}}}}
$$

Demanding $ u = 0 $ for $ x = \pm (1 \mp {\eta}) $, gives us,

\be
{\Gamma\big ({3\over4} - {i\chi\over 2} - {iEl\over 2}\big )
\Gamma\big ({3\over4} + {i\chi\over 2} - {iEl\over 2}\big )
\over \Gamma\big (1 - iEl\big )}\big ({1\over 2\eta}\big )^{iEl\over 2}
- c.c = 0, 
\ee
where $c.c$ denotes the complex conjugate of the first term. 
Here the two gamma funtions are complex conjugate to each other 
and the phase factor involves $\chi$ and $E$. With
proper choice of the normalization constant, we have,
\be
{e^{i[{El\over2}{\ln({1\over{2\eta}})} + {\pi}{{\alpha_1}(E,\chi)}]}} - {c.c} =
0.
\ee

Here $\alpha_1$ is a function of $E$ and $\chi$ 
such that ${0 \leq |\alpha_1|<1}$.
Hence we have, from equ.(20),
\be
{{El\over2}{\ln({1\over{2\eta}})} + {\pi}{\alpha_1}} = {{n_1}{\pi}}
\ee
where $n_1$ is an integer. Now in the above relation as $\eta \rightarrow 0$,
both the r.h.s, and the first term in the l.h.s become very large compared
to the second term in the l.h.s. Hence we have
\be
E = {{2{n_1}\pi}\over {l[\ln({1\over{2\eta}})]}}.
\ee
Clearly the solution diverges as $\eta \rightarrow 0$.

Proceeding in the same way for the second sector (equ.(12))
of $\nu$ we obtain another set of values
of E :
\be
{{{{\Gamma(1 - {\lambda\over2} - {iEl\over2})}
{\Gamma({1\over2} + {\lambda\over2} - {iEl\over2})}}\over
{\Gamma(1 - iEl)}}}{{({1\over2\eta})}^{iEl\over2}} - c.c = 0
\ee
This condition gives us another set of values of E:
\be
{{El\over2}{[\ln({1\over{2\eta}})]}} + {\pi{\alpha_2}(E,\lambda)} = {n_2}{\pi} 
\ee
where $n_2$ is an integer and ${0\leq |\alpha_2|<1}$.
Now proceeding in the same way as
the first case we have,
\be
E = {{2{n_2}\pi}\over {l[\ln({1\over{2\eta}})]}}.
\ee
Clearly this solution also diverges for $\eta \rightarrow 0$.

The free energy of the scalar field is given by
\be
{{\beta}F} = {{\sum_{\rm degeneracies}}}{{\sum_{n}}}
{\ln{(1 - {e^{-{{\beta}{E_n}}}})}}
\ee
In this case we have two sectors of $\nu$ and correspondingly,
two sets of values of the energy levels. Hence the free energy
of the scalar field is given by the following expression,
\ba
{{\beta}F} = &{{1\over\pi}}&
{\sum_N}{\int_{0}^{1}d\lambda}
{\int_{r_h + \epsilon}^{L_1}dr~k(r,\lambda,N)}{\sum_{n_1}}
{\ln{(1 -{e^{-{{\beta}{E_{n_1}}}}})}} \nonumber\\   
&+&
 {1\over\pi}{\sum_N}{\int_{-\chi_1}^{\chi_1}d\chi}
{\int_{r_h + \epsilon}^{L_2}dr~k(r,\chi,N)}{\sum_{n_2}}
{\ln{(1 -{e^{-{{\beta}{E_{n_2}}}}})}}
\ea
Here $\epsilon$ is the radial brick wall cut-off
parameter. It is evident from equ.(14) and equ.(15) that
the respective expressions for $k(r,\chi,N)$ and $k(r,\lambda,N)$
are divergent as the brick wall is pulled on to the 
horizon (i.e, as $\epsilon\rightarrow 0$). The $\chi$
integration is cut-off at some large but finite absolute
value $\chi_1$ of $\chi$.

For the degeneracy factors associated with energy, we have from equ. (23),
and equ. (24)
\be
{dE} = {{2\pi}\over{l}}{dn\over{\ln{({1\over{2\eta}})}}}
\ee
For the radial degeneracy factors we have from equ.(14) and equ.(15):
\ba
&&g (\epsilon,L_1,\lambda,N)  =\nonumber\\ 
&&{1\over{2\pi{r_h}^2}}{\big[{
\sqrt{{\lambda(1 - \lambda){r^2}{r_h}^2} - {{N^2}{l^2}({r^2 - {r_h}^2})}
- {{m^2}{l^2}{r^2}({r^2 - {r_h}^2})}}
{\ln{\big({{r^2 - {{r_h}^2}}\over{r^2}}\big)}}\big]}^{L_1}_{r_h}} -
\nonumber\\
&&{1\over{4\pi{r_h}^2}}{\int_{r_h + \epsilon}^{L_1}
{\ln{\big({{r^2 - {{r_h}^2}}\over{r^2}}\big)}}
{{{2\lambda(1 - \lambda)r{r_h}^2} - {2{N^2}{l^2}r - {4{m^2}{l^2}{r^3}}
+ {{2m^2}{l^2}r{r_h}^2}}\over{\sqrt{{\lambda(1 - \lambda){r^2}{r_h}^2} -
{{N^2}{l^2}({r^2 - {r_h}^2})}
- {{m^2}{l^2}{r^2}({r^2 - {r_h}^2})}}}}dr}}
\ea
and a similar expression for $g(\epsilon,L_2,\chi,N)$ with $\lambda(1 - \lambda)$
replaced by $({{\chi}^2 + {1\over4}})$ and $L_1$ replaced by $L_2$
in the above equation. From these expressions we have the 
following expression for the free energy of the scalar field,
\ba
{\beta}F =
&{{l\over{2\pi}}}& {\sum_N}
{\big[\int_{0}^{1}d\lambda{g(\epsilon,L_1,\lambda,N)}\big]}
{\big[{\ln({1\over{2\eta}})}{{\int}{dE}}{\ln{(1 -
{e^{-{{\beta}{E}}}})}}\big]}
\nonumber\\
&+& {l\over{2\pi}}{\sum_N}
{\big[\int_{0}^{1}d\lambda{g(\epsilon,L_2,\chi,N)}\big]}
{\big[{\ln({1\over{2\eta}})}{{\int}{dE}}{\ln{(1 -             
{e^{-{{\beta}{E}}}})}}\big]}
\ea

Now we are interested 
in the divergence of the field modes near the horizon, i.e,
the divergence appearing for small values of $\epsilon$.
This gives the usual ultraviolet divergence of the free
energy of the scalar field similar the case of Schwarzschild
or Reissner-Nordstrom [3,5] black holes. Hence confining our
attention to the near horizon region we have the following
expression for the free energy:
\ba
{\beta}F =
&{l\over{2\pi}}&{\sum_N}
{\big[\int_{0}^{1}d\lambda{g(\epsilon,\lambda,N)}\big]}
{\big[{\ln({1\over{2\eta}})}{{\int}{dE}}{\ln{(1 -
{e^{-{{\beta}{E}}}})}}\big]}   
\nonumber\\
&+& {l\over{2\pi}}{\sum_N}
{\big[\int_{0}^{1}d\lambda{g(\epsilon,\chi,N)}\big]}
{\big[{\ln({1\over{2\eta}})}{{\int}{dE}}{\ln{(1 -
{e^{-{{\beta}{E}}}})}}\big]}
\ea
where $g(\epsilon,\lambda,N)$ and $g(\epsilon,\chi,N)$,the
ultraviolet divergent parts of the radial degeneracy factors
are given by:
\be
g(\epsilon,\lambda,N) = -{{1\over2\pi{r^{2}_{h}}}
{\ln{({2\epsilon\over{r_{h}}})}}
{{\big[{{{r^{4}_{h}}\lambda(1 - \lambda)} - 
{2\epsilon{r_h}{N^2}{l^2}} - 
{2\epsilon{r^{3}_{h}}{m^2}{l^2}}}\big]}^{1\over2}}} 
\ee
and
\be
g(\epsilon,\chi,N) = -{{1\over2\pi{r^{2}_{h}}}
{\ln{({2\epsilon\over{r_{h}}})}}
{{\big[{{{r^{4}_{h}}({\chi^2 + {1\over4}})} - 
{2\epsilon{r_h}{N^2}{l^2}} - 
{2\epsilon{r^{3}_{h}}{m^2}{l^2}}}\big]}^{1\over2}}}
\ee

Hence to the leading order in $\epsilon$ we have the 
following expression for the free energy of the 
scalar field:
\be
F = {{N_1}l\over{32\pi{\beta}^2}}{\big[{{\pi} +
{4{\chi_1}\sqrt{4{\chi^{2}_{1}} + 1}}}
+ {4\ln({\sqrt{4{\chi^{2}_{1} + 1}} -
2\chi})}\big]}{\big[{\ln({1\over{2\eta}})}\big]}
{\big[{\ln({{2\epsilon}\over{r_h}})}\big]}
\ee
where the $N$ integration is cut-off at some large
but finite values $N_1$. The free energy is clearly
divergent as the brick wall is pulled back on the horizon with
an additional logarithmic divergence in the $\theta$- cut-off
$\eta$.

The entropy is given by
$$
S = {{{\beta}^{2}}{dF\over{d{\beta}}}}
$$
This gives
\be
S =  {{N_1}l\over{16\pi{\beta}}}{\big[{{\pi} +
{4{\chi_1}\sqrt{4{\chi^{2}_{1}} + 1}}}
+ {4\ln({\sqrt{4{\chi^{2}_{1} + 1}} -
2\chi})}\big]}{\big[{\ln({1\over{2\eta}})}\big]}
{\big[{\ln({{r_h}\over{2\epsilon}})}\big]}
\ee
Now we take the inverse temperature to be
the period of the time coordinate in the
Euclidean sector of the metric to avoid the conical singularity.
In this case it is ${2{\pi}l}$.
Substituting in equ.(36), we have
\be
S = {{N_1}\over{32\pi^2}}{\big[{{\pi} +
{4{\chi_1}\sqrt{4{\chi^{2}_{1}} + 1}}}
+ {4\ln({\sqrt{4{\chi^{2}_{1} + 1}} - 
2\chi})}\big]}{\big[{\ln({1\over{2\eta}})}\big]}
{\big[{\ln({{r_h}\over{2\epsilon}})}\big]}
\ee
Thus the entropy is not proportional to the horizon surface
area which, in this case, is given by ${2\pi{r_h}}$. However,
it is logarithmically divergent in both the radial coordinate $r$
and angular coordinate $\theta$.

\section{{Discussion}}
In conclusion, we have considered the thermodynamics of a scalar
field in 3+1 -dimensional CCBH background.
In order to regulate the solution of the scalar field wave
equation we have introduced two cut-off parameters. The first
one is the usual radial (brick wall) cut-off parameter $\epsilon$
which regulates the solution near the horizon and the second one
is the cut-off parameter $\eta$ in the angular coordinate
$\theta$.
We have calculated the entropy
using the WKB approximation and found it
to be logarithmically divergent in both of these cut-off parameters.

The expression for entropy
is remarkable. Firstly, it is not proportional to the area $({2{\pi}{r_h}})$ of
the horizon.
However, this behaviour is also reflected in the
expression for the gravitational entropy of the 3+1 -dimensional CCBH which
is given by [11]:
$$
{S_{grav}} = {{\pi}{l^{2}}{f(R)}},
$$
where $R$ is the radius of the quasilocal surface used to foliate
the space-time. The gravitational entropy of 4+1 -dimensional rotating CCBH
is also not proportional to the horizon surface area. So it seems
to be a general feature associated with  constant curvature
black holes.

Secondly, the divergence of the scalar field entropy is closely
related to the boundary conditions. The first, expected, diverging
term, ${\ln({{\epsilon}\over{r_h}})}$, is related to the brick wall boundary
condition near the horizon. However the second diverging
factor  ${\ln({2\over{\eta}})}$  is  surprising  because it is not
seen in scalar
field entropy in conventional black hole backgrounds (Schwarzschild or
Reissner - Nordstrom). The reason for this factor
is exactly the same as that which is
responsible for the peculiarity of the gravitational entropy
in 3+1 -dimensional CCBH, i.e, the additional degeneracy of
the foliation used. In the case of 4+1 -dimensional CCBH the metric has the
same form as the 3+1 -dimensional CCBH; the only difference is that
the surfaces of constant $r$ and $t$ have the
topology ${{S_{2}}\otimes{S_{1}}}$. It is expected that to regularize
the solution of the  scalar field wave equation a similar angular cut-off
parameter will be necessary.

It may be noted that the study of scalar fields in BTZ
black hole background [15 , 16] has shown that the entropy
of scalar fields depends on
the method of calculation used. In this context it will be
interesting to calculate the scalar field entropy in 3+1 -dimensional
CCBH background using other techniques (e.g, heat kernal expansion
or Hartle - Hawking Green function).

\section{{Acknowledgments}}
It is a great pleasure to the author to thank 
Prof. P. Mitra for many helpful
discussions.

\end{document}